\documentclass[a4paper,fleqn]{cas-dc}

\usepackage[numbers]{natbib}

\def\tsc#1{\csdef{#1}{\textsc{\lowercase{#1}}\xspace}}
\tsc{WGM}
\tsc{QE}
\tsc{EP}
\tsc{PMS}
\tsc{BEC}
\tsc{DE}
\begin{document}
\let\WriteBookmarks\relax
\def\floatpagepagefraction{1}
\def\textpagefraction{.001}
\shorttitle{Rivalry in micro-convection}
\shortauthors{G.Kitenbergs et~al.}

\title [mode = title]{Rivalry of diffusion, external field and gravity in micro-convection of magnetic colloids}                      
%\tnotemark[1,2]

%\tnotetext[1]{This document is the results of the research  project funded by the National Science Foundation.}

%\tnotetext[2]{The second title footnote which is a longer text matter   to fill through the whole text width and overflow into   another line in the footnotes area of the first page.}

\author[1]{Guntars Kitenbergs}[type=editor,
                        orcid=0000-0002-2230-713X]
\cormark[1]
%\fnmark[1]
\ead{guntars.kitenbergs@lu.lv}
%\ead[url]{mmml.lu.lv}

%\credit{Conceptualization of this study, Methodology, Software}

\address[1]{MMML lab, University of Latvia, Jelgavas 3, Riga, LV-1004, Latvia}

\author[1]{Andrejs C\={e}bers}

%\credit{Data curation, Writing - Original draft preparation}

\cortext[cor1]{Corresponding author}
%\cortext[cor2]{Principal corresponding author}
%\fntext[fn1]{This is the first author footnote. but is common to third author as well.}
%\fntext[fn2]{Another author footnote, this is a very long footnote and  it should be a really long footnote. But this footnote is not yet  sufficiently long enough to make two lines of footnote text.}

%\nonumnote{This note has no numbers. In this work we demonstrate $a_b$  the formation Y\_1 of a new type of polariton on the interface  between a cuprous oxide slab and a polystyrene micro-sphere placed  on the slab.  }

\begin{abstract}
Magnetic fields and magnetic materials have promising microfluidic applications. 
For example,  magnetic micro-convection can enhance mixing considerably.
However, previous studies have not explained increased effective diffusion during this phenomenon.
Here we show that enhanced interface smearing comes from a gravity induced convective motion within a thin microfluidic channel, caused by a small density difference between miscible magnetic and non-magnetic fluids.
This motion resembles diffusive behavior and can be described with an effective diffusion coefficient.
We explain this with a theoretical model, based on a dimensionless gravitational Rayleigh number, and verify it by numerical simulations and experiments with different cell thicknesses.
Results indicate the applicability and limitations for microfluidic applications of other colloidal systems.
Residual magnetic micro-convection follows earlier predictions.

\end{abstract}

%\begin{graphicalabstract}
%\includegraphics{figs/grabs.pdf}
%\end{graphicalabstract}

%\begin{highlights}
%\item Research highlights item 1
%\item Research highlights item 2
%\item Research highlights item 3
%\end{highlights}


\maketitle

\section{Introduction}
Concepts of microfluidics and lab--on--a--chip systems are attractive for various biological and medical problems, however, simple and effective solutions are still needed to overcome typical limitations and enable applications \cite{MFintro}.
Diffusion limited mixing of fluids in the laminar microfluidics flow is among those.
Use of magnetism and magnetic materials offer multiple effective and simple mechanisms to improve mixing, as has been shown in recent reviews \cite{ReviewMixMagMat,MagNanoMix,MicroMagnFluid}.

\begin{figure}
	\centering
		\includegraphics[width=\columnwidth]{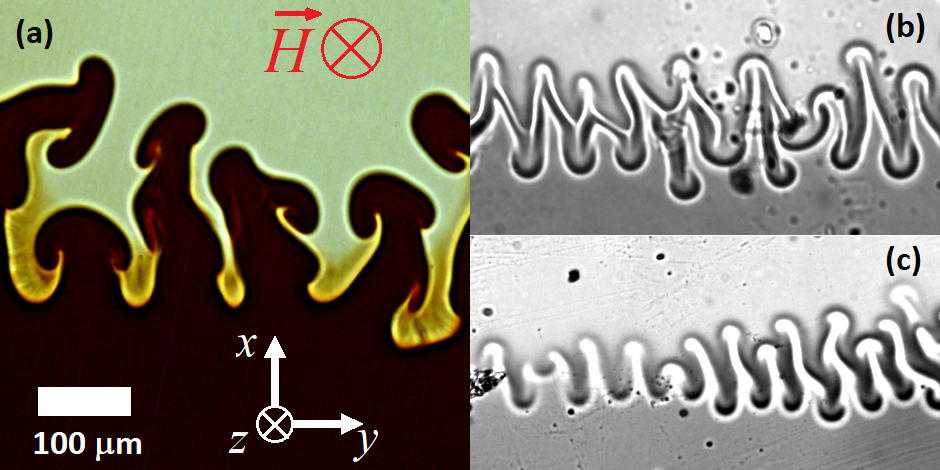}
	\caption{Characteristic fingering pattern of the magnetic micro--convection phenomena. Finger size $\lambda$ approximately agrees with the cell thickness $h$. (a) $\lambda\approx h=130~\mu$m, (b) $\lambda\approx h=50~\mu$m, (c) $h=25~\mu$m and $\lambda\approx35~\mu$m.}
	\label{FIG:Pattern}
\end{figure}

A particularly interesting type of magnetic micromixers are based on a phenomenon called magnetic micro--convection, discovered by \citet{MCV-begin}.
It is a convective fingering pattern (for example, see Fig.\ref{FIG:Pattern}) that emerges on an interface of miscible magnetic and nonmagnetic fluids when exposed to an external magnetic field, perpendicular to the fluid plane.
This comes from the rivalry between diffusion and self-magnetic field of magnetic fluid.
Due to complex nature and simple implementation, this phenomenon has been widely studied for various conditions and applications, including a flat cell \cite{Derec2008,JFM}, radial geometry \cite{ChenRadial,WenRadial}, an interplay with Rosensweig instability \cite{hybrid}, microfluidic mixing \cite{JMMM} and surface patterning with sessile drops \cite{Sessile}, both experimentally and theoretically.
In \cite{JFM2} we have updated the theoretical model to an extent that it quantitatively describes the experimental observations.
However, for this comparison, an effective diffusion coefficient $D_{\text{eff}}$ was used instead of the experimentally measured diffusion coefficient of magnetic nanoparticles $D$.
Moreover, $D_{\text{eff}}$ was estimated to be two order of magnitude larger than $D$.
In this paper we investigate the reasons behind this extraordinary situation.

It turns out that a small density difference between miscible fluids can be of importance even in microchannels.
If a magnetic micro-convection experiment is performed in a system, which is turned sideways and where the denser magnetic fluid is below the less dense nonmagnetic fluid, one can observe a normal diffusion with a coefficient $D$ \cite{EPJE}.
Here we start with revisiting a model for gravity induced concentration smearing on the interface, as proposed in \cite{JFM}.
Using numerical simulations we show how it causes a density difference induced convective motion within the thickness of the channel.
We perform experiments in thinner channels and compare numerical and experimental results, using an effective diffusion coefficient $D_{\text{eff}}$.

\section{Models, materials and methods}

\subsection{Theoretical model for the magnetic micro--convection and interface diffusion}
\label{sec:mcv}
For the magnetic micro--convection we consider two miscible magnetic and non--magnetic fluids, which are confined in a horizontal Hele-Shaw cell and exposed to a homogeneous magnetic field, which is perpendicular to the cell.
At time $t=0$ we assume a straight interface with a concentration step, where $c=1$ corresponds to magnetic fluid with maximal magnetic nanoparticle concentration and $c=0$ to non--magnetic fluid. 
The process can be theoretically modelled with a system of the Brinkman, continuity and convection-diffusion equations, as described in \cite{JFM2}.
Without going in details, we note that the phenomenon is characterized with a dimensionless magnetic Rayleigh number $Ra_m=\frac{(\chi H)^2h^2}{12\eta D}$, where $\chi$ is the susceptibility of magnetic fluid, $H$ is the magnetic field, $h$ is the cell thickness, $\eta$ is the fluid viscosity, assumed to be equal across the fluid, and $D$ is the diffusion coefficient of magnetic nanoparticles.
A critical field $H_c$ is necessary for the instability to appear. 
This corresponds to a critical magnetic Rayleigh number $Ra_m^{\text{crit}}\approx 6$, while the finger size $\lambda$ is approximately equal to the cell thickness $h$ and does not depend on $Ra_m$ \cite{JFM2}.

If no magnetic field is applied, the magnetic and non--magnetic fluids should slowly mix via diffusion.
This can be described by the Fick's law.
For the case of an initial concentration step, the solution is $c(x,t)=(c_0/2)\cdot[1-\text{erf}(x/(2\sqrt{Dt}))]$, where $\text{erf}$ is the Gaussian error function and $c_0$ is the initial concentration.
Concentration profiles $c(x)$ can be measured experimentally and used to calculate a diffusion coefficient.
For easier comparison between different concentration profiles, we use concentration gradient $\partial{c}/\partial{x}$ at the initial interface $x=0$.
One can show that $\partial{c}/\partial{x}$ is linked to the diffusion coefficient $D$ via relation
\begin{equation}
\label{eq:dcdx}
    \left(4\pi\left(\frac{\partial{c}}{\partial{x}}\right)^2\right)=D t.
\end{equation}

\subsection{Theoretical model for gravitational influence}
\label{sec:gravity}
A model from \cite{JFM} characterizes the gravitational influence on a miscible fluid interface when there is a density difference between the fluids.
The resulting effect in the $x-z$ plane (see the definition of axis in Fig.\ref{FIG:Pattern}) is described by the Stokes model with a concentration $c$ dependent gravity force and the diffusion equation.
The corresponding set of partial differential equations (PDEs) in dimensionless form is
\begin{eqnarray}
-\nabla p+\Delta \vec{v}-c\vec{e_{z}}&=&0 \label{eq:2}\\
\frac{\partial c}{\partial t} +Ra_g(\vec{v}\cdot \nabla)c &=&\Delta c,\label{eq:3}
\end{eqnarray}
where $Ra_g=\Delta\rho gh^{3}/8D\eta$ is the gravitational Rayleigh number.
It is obtained by scaling time by $h^{2}/4D$, length by $h/2$ and the velocity by $\Delta\rho gh^{2}/4\eta$, where $\Delta\rho$ is the density difference, $g=981$~cm$\cdot$s$^{-2}$ is standard gravity and $h$, $D$ and $\eta$ as introduced previously.

\subsection{Numerical simulations}
\label{sec:sim}
Numerical simulations are performed in COMSOL, following \cite{JFM}.
The simulation is defined with PDEs (\ref{eq:2},~\ref{eq:3}) for a two dimensional side view of a cell in the $x-z$ plane with a slightly smeared normalized step-like concentration interface (closer to experiments). 
No--slip boundary condition is used.
The cell size is defined in dimensionless units with a thickness $2$ and the width $30$ for $Ra_g>1000$ and $10$ for smaller $Ra_g$.
Solutions are searched from times $t=0..1$ with a $\Delta t=0.001$ interval for multiple gravitational Rayleigh number $Ra_g$ values.

To quantify numerical simulation results and compare them with experimental observations, it is useful to average the concentration along the thickness of the cell.
This gives the same information as in a microscopy image.
We arrive at an average concentration profile along $x$ axis $\bar{c}(x)=\frac{1}{2}\int_{-1}^{1} c(x,z) dz$, where factor $1/2$ comes from the cell thickness, which is $2$. 
This step can be directly implemented in COMSOL, using \textit{linproj1} operator.

\subsection{Microscopy and microfluidics}
\label{sec:mf}
To observe the phenomena, we use an inverted microscope Leica DMI3000B. 
It is equipped with a pair of coils and can create a homogeneous magnetic field $H=0..150$~Oe, which is perpendicular to the plane of observation (along $z$ axis in Fig.\ref{FIG:Pattern}).
Images are taken with microscope cameras (Basler acA1920-155um, Lumenera Lu165C) and $10\text{x}$ or $40\text{x}$ objectives.
Image intensities are related to concentration fields via Beer-Lambert law. 
To improve the extracted data quality from images, especially for thinner cells, manual image processing is done, including filtering, masking areas with dirt, averaging, etc.

In experiments we use two fluids.
Water based magnetic fluid is made by a co-precipitation method that forms maghemite nanoparticles, which are stabilized with citrate ions.
The resulting nanoparticle suspension has a volume fraction $\Phi=2.8\%$, average diameter $d=7.0$~nm, saturation magnetization $M_{sat}=8.4$~G and magnetic susceptibility $\chi_m=0.016$, as determined by a vibrating sample magnetometer (Lake Shore 7404).
Both Dynamic Light Scattering (Malvern NanoZS) and Force Rayleigh Scattering (setup at PHENIX lab, Sorbonne University) methods give particle diffusion coefficient $D\approx2.5\cdot10^{-7}$~cm$^2\cdot$s$^{-1}$.
For non--magnetic fluid we take distilled water.
We assume both viscosities to be equal to that of water $\eta=0.01$~P.
The density difference between the two fluids is $\Delta\rho_0=0.148$~g/cm$^3$.

To bring fluids to a sharp interface in a flat cell, we use microfluidics chips with a $Y$ channel shape.
We fabricate chips with 3 different thicknesses - $h=130~\mu$m, $h=50~\mu$m and $h=25~\mu$m.
Fluids are driven through two inlets at a flowrate $Q$ by a syringe pump PHD Ultra from Harvard Apparatus and via FEP tubing (IDEX).
We approximate the fluid velocity in the channel as $v=Q/(w\cdot h)$, where $w$ is the channel width.
The chip has a single outlet, which is left open (zero pressure).

The thickest chip $h=130~\mu$m is made by welding a precut Parafilm spacer of the desired channel shape between two glass slides.
Holes are drilled in one of the glass slides to glue in tubing connectors.
For micro-convection experiments a channel shape that allows to merge two droplets is used, as explained in \cite{JFM2}.
For no--field experiments we exploit a continuous microfluidics channel, as described in \cite{EPJE}.
In this way it is possible to quickly obtain channels of the desired configuration with a width down to $w\approx1$~mm (for example, see Fig.\ref{FIG:MFdiff}(a)).

Two thinner cells ($h=50~\mu$m and $h=25~\mu$m) are made by following the rapid prototyping routine \cite{PDMS}.
We use molds with SU8 photoresist features on a glass substrate, fabricated in the Institute of Solid State Physics of the University of Latvia.
PDMS (Sylgard 184, Dow corning) is mixed and then cured in an oven for $4$ hours at $65^\circ$C.
After removing PDMS from mold, holes for tubing connections are made.
Then surface of PDMS is treated with a Corona SB plasma cleaner (BlackHoleLab). 
The same is done for a $24\times24$~mm$^2$ $0.19$~mm thin microscope glass slide.
After treatment both pieces are put together and left for a few hours.
The resulting chip has two $200~\mu$m wide channels merging into one $w=400~\mu$m wide channel (for example, see Fig.\ref{FIG:MFdiff}(b)\,\&\,(c)).

\section{Results and discussion}

In previous experiments \cite{JFM2}, where we observed the characteristic fingering pattern of the magnetic micro--convec-tion (Fig.\ref{FIG:Pattern}(a)), the cell had a thickness $h=130~\mu$m and the fluid density difference was $\Delta\rho_0=0.148$~g/cm$^3$.
This corresponds to a gravitational Rayleigh number $Ra_g=13'500$, according to the model of gravitational influence, described in \S\ref{sec:gravity}.
It is much larger than $1$ and suggests a significant gravitational effect.

\begin{figure*}
	\centering
		\includegraphics{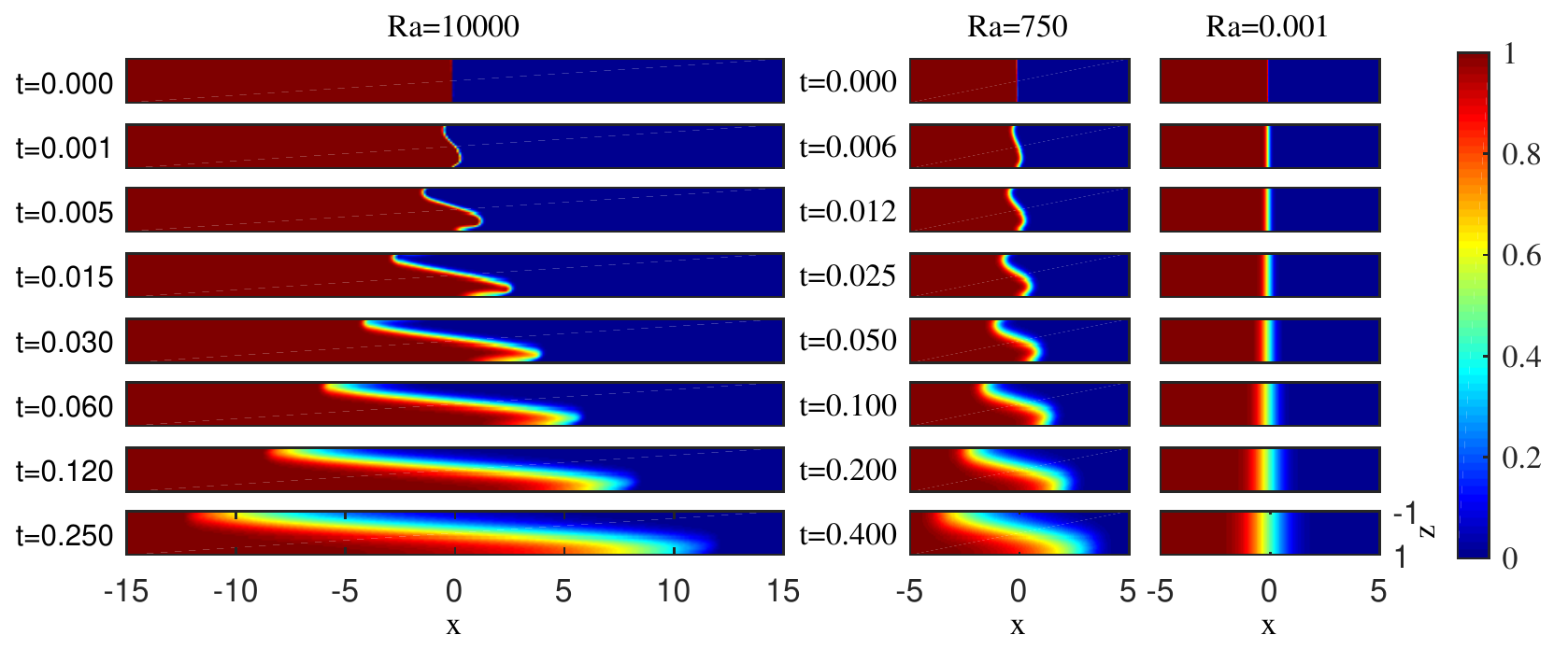}
	\caption{Numerical simulation results of concentration field dynamics as viewed from a side of the cell ($x-z$ plane) for three different gravitational Rayleigh numbers $Ra_g$ $10000$, $750$ and $0.001$. Decrease in $Ra_g$ terminates convective motion, leaving only diffusion. Red corresponds to initial magnetic fluid, while blue - non--magnetic fluid.}
	\label{FIG:sim}
\end{figure*}

To investigate this in detail, we perform numerical simulations of the dimensionless model for the case of no magnetic field.
We find the concentration plot dynamics of $x-z$ plane (side--view of the cell) for a variety of $Ra_g$, ranging from $10^{-3}$ to $2\cdot10^{4}$.
Characteristic results can be seen in Fig.\ref{FIG:sim}.
For a large $Ra_g$ (e.g. $Ra_g=10'000$ in Fig.\ref{FIG:sim}), the denser magnetic fluid (red) quickly slides underneath the less dense non--magnetic fluid (blue).
Eventually the diffusive mixing takes over and smears the deformed interface.
Also for a medium $Ra_g$ (e.g. $Ra_g=750$ in Fig.\ref{FIG:sim}), the denser magnetic fluid quickly slides underneath the less dense non--magnetic fluid, however, the deformed interface is much smaller and diffusion takes over faster.
For small $Ra_g$ (e.g. $Ra_g=0.001$ in Fig.\ref{FIG:sim}) no interface deformation can be seen and diffusion slowly mixes both fluids.

\begin{figure*}
	\centering
		\includegraphics{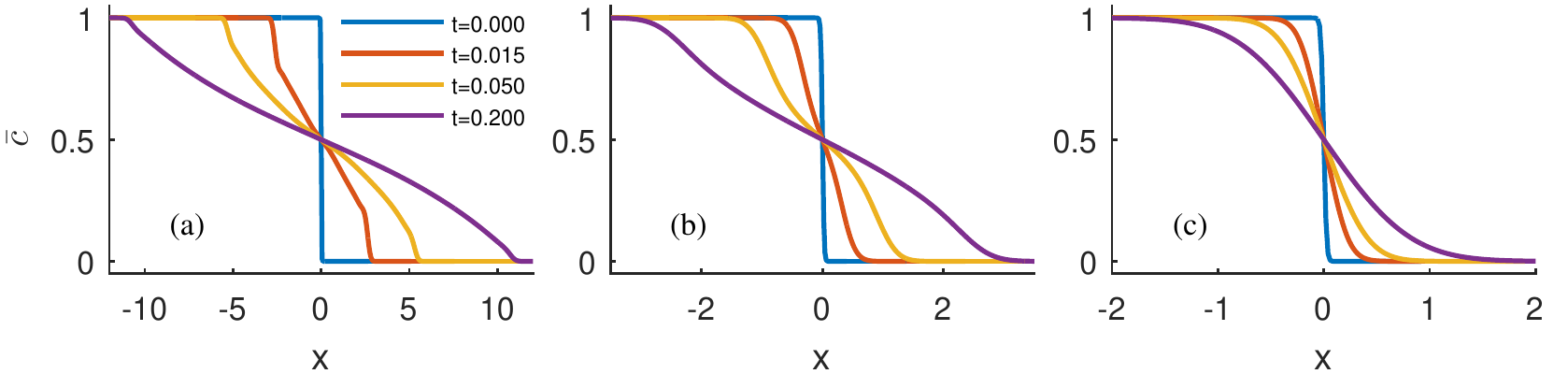}
	\caption{Average concentration $\bar{c}$ profiles at four different times $t$ for numerical simulation results of three different gravitational Rayleigh numbers (a) $Ra_g=10000$, (b) $Ra_g=750$ and (c) $Ra_g=0.001$.}
	\label{FIG:cplots_sim}
\end{figure*}

Due to limitations of the experimental system, it is impossible to observe concentration field dynamics in $x-z$ plane directly.
For comparison, as explained in \S\ref{sec:sim}, it is worth to calculate the average concentration profiles $\bar{c}(x,t)$.
Examples for such profiles are given in Fig.\ref{FIG:cplots_sim}.
Similar profiles can be seen in all cases.
A notable difference is only visible close to $c=0$ and $c=1$.
For large $Ra_g$ the transition to the non-mixed areas is sharp (Fig.\ref{FIG:cplots_sim}(a)), while for small $Ra_g$ the transition is smooth (Fig.\ref{FIG:cplots_sim}(c)).
These differences can be used to identify the convective motion within the cell.

\begin{figure}
	\centering
		\includegraphics[width=\columnwidth]{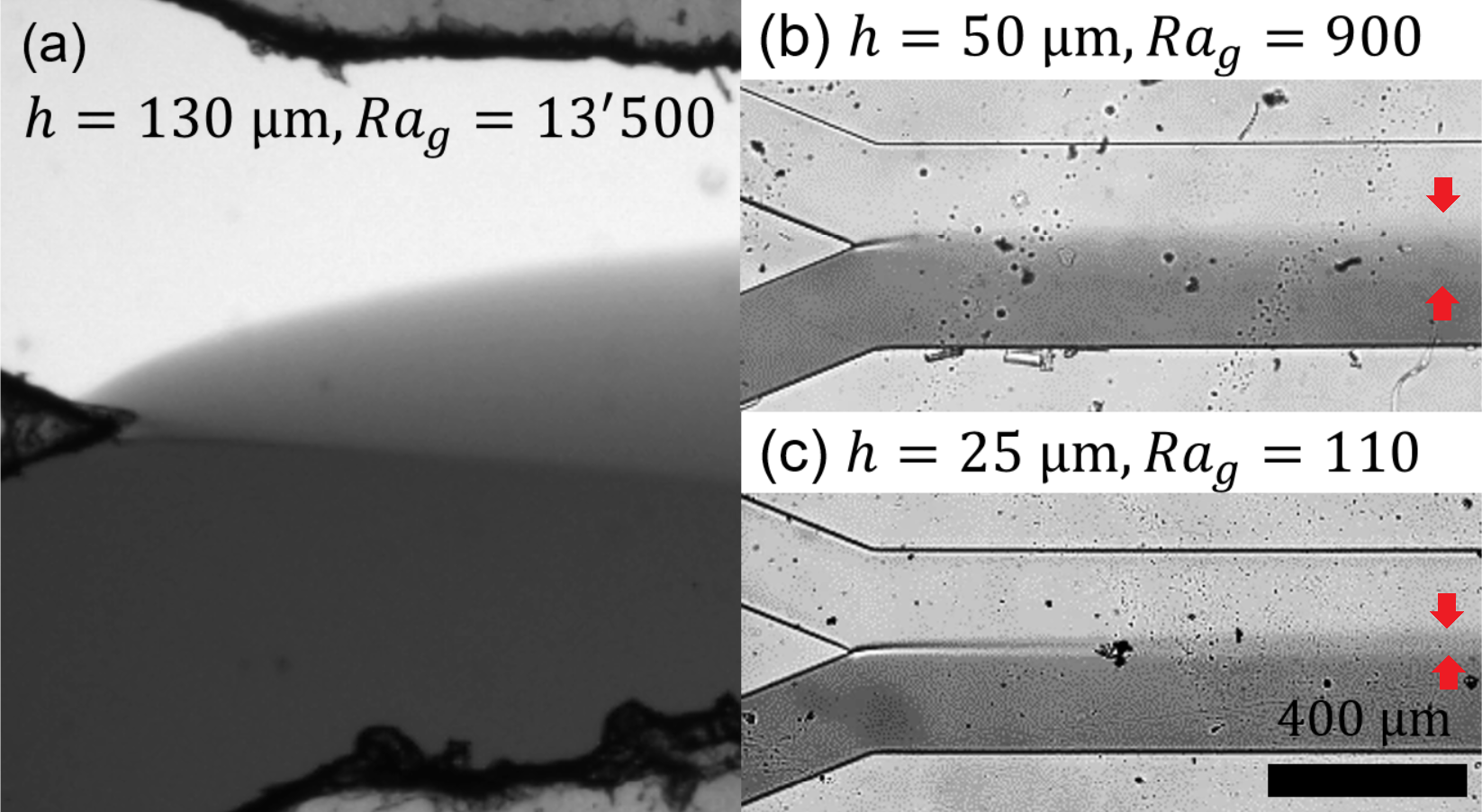}
	\caption{Terminating gravity induced convective motion by reducing the channel thickness. (a)$h=130~\mu$m, $v\approx440~\mu$m/s, (b)$h=50~\mu$m, $v\approx333~\mu$m/s, (c)$h=25~\mu$m, $v\approx333~\mu$m/s. Red arrows identify interface smearing.}
	\label{FIG:MFdiff}
\end{figure}

From the definition of gravitational Rayleigh number $Ra_g$, it is clear that the gravitational influence for this system can be decreased by reducing the thickness of the cell $h$.
For experiments we use microfluidics cells with three different thicknesses, as described in \S\ref{sec:mf}.
For the same fluid pair, the corresponding gravitational Rayleigh numbers are $Ra_g=13'500$, $Ra_g=900$ and $Ra_g=110$.
Measurements are done in continuous microfluidics regime, where both magnetic and nonmagnetic fluids are brought to a contact and the change in the interface is observed along the microfluidic channel.
Sample images with the same magnification are given in Fig.\ref{FIG:MFdiff}.
Fluids flow from the left to right.

Flowrates $Q$ are chosen so that the fluid velocity in channels is similar.
For the thickest channel $h=130~\mu$m it is $Q=4.8~\mu$L/min, which gives $v\approx440~\mu$m/s.
A clear smearing of the interface is visible, reaching around $400~\mu$m by the end of the field of view (Fig.\ref{FIG:MFdiff}(a)).
For thinner channels $h=50~\mu$m and $h=25~\mu$m lower flowrates $Q=0.4~\mu$L/min and $Q=0.2~\mu$L/min are used, which give the same velocity $v\approx333~\mu$m/s and a much smaller smearing is visible.
In Fig.\ref{FIG:MFdiff}(b) with $h=50~\mu$m it reaches $\approx150~\mu$m, while for with $h=25~\mu$m in Fig.\ref{FIG:MFdiff}(c) smearing is only $\approx50~\mu$m.
Qualitatively we observe that the gravity induced convection is terminated.

\begin{figure*}
	\centering
		\includegraphics{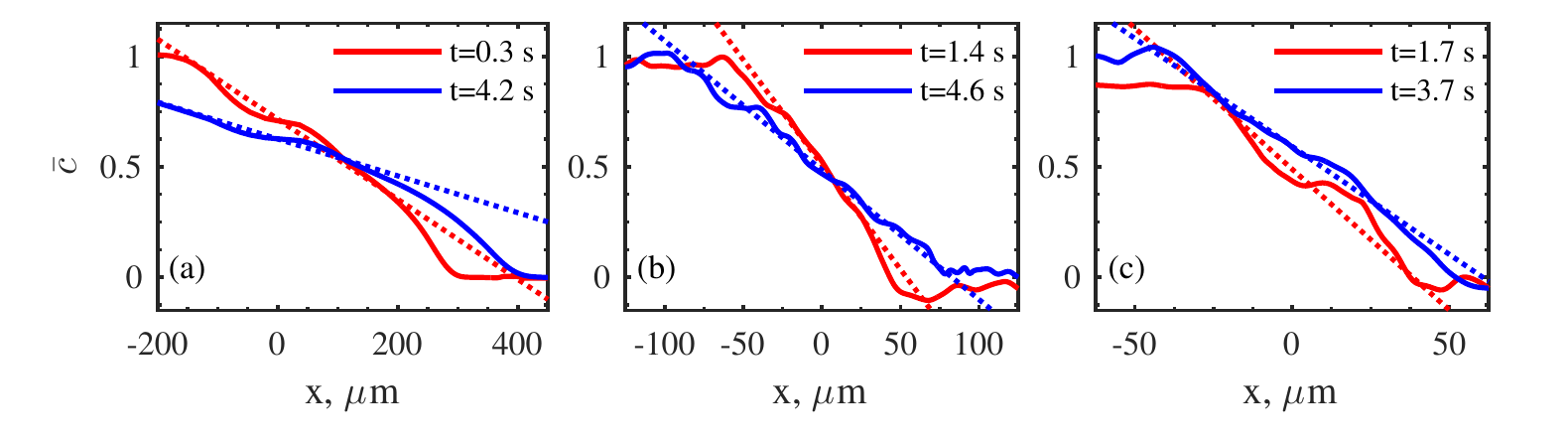}
		\includegraphics{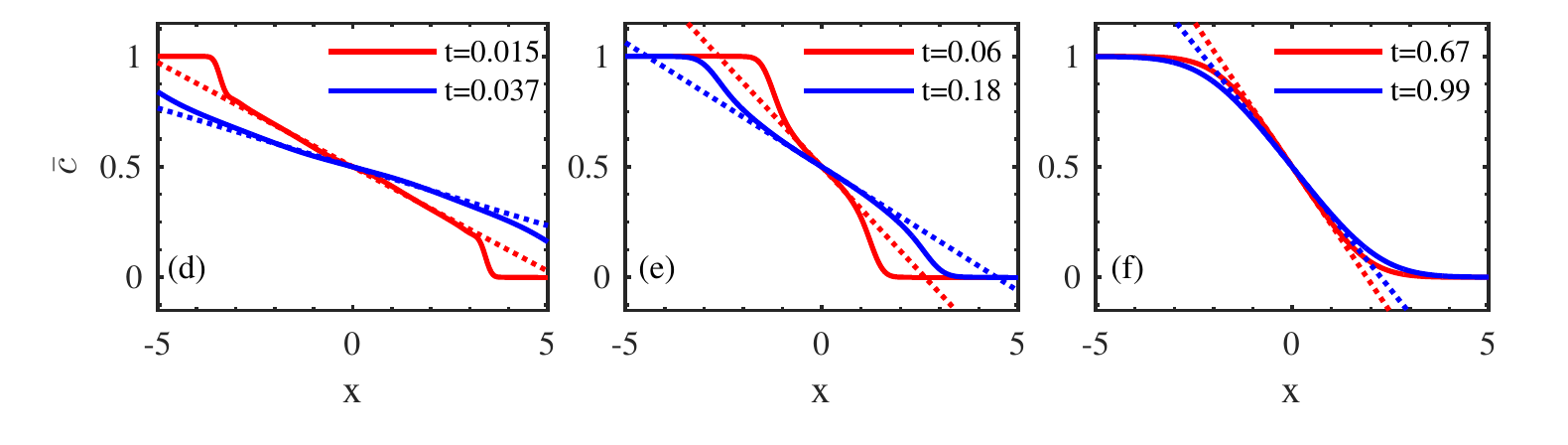}
	\caption{Average concentration $\bar{c}(x,t)$ dynamics. Experimental results for (a) $h=130~\mu$m ($Ra_g\approx13500$), (b) $h=50~\mu$m ($Ra_g\approx900$) and (c) $h=25~\mu$m ($Ra_g\approx110$). Numerical simulation results for (d) $Ra_g=15000$, (e) $Ra_g=1000$ and (f) $Ra_g=100$. Dotted lines are fits of concentration gradient $\delta c/\delta x$ near $x=0$. Scale of x-axis and difference between two $t$ in experimental results (a)-(c) have been chosen to fit with corresponding dimensionless quantities of numerical results in (d)-(e).}
	\label{FIG:ConcPlots}
\end{figure*}

From images in Fig.\ref{FIG:MFdiff} we obtain average concentration $\bar{c}(x,t)$ dynamics in experiments.
Concentrations are found from intensity maps via Lambert-Beer law. 
Assuming a constant fluid velocity $v$, translation along $y$-axis can be converted to time $t=y/v$.
Experimental average concentrations $\bar{c}$ for each of the thicknesses at two different times $t$ are shown in Fig.\ref{FIG:ConcPlots}(a)-(c).
For the thickest cell (Fig.\ref{FIG:ConcPlots}(a)) a similar sharp transition of concentration profile near non-mixed areas as in Fig.\ref{FIG:cplots_sim}(a) can be seen, confirming that the smearing is induced by gravity effects.
It is visible that the concentration profile is not symmetric and a slight bump can be seen for concentrations just above $c=0.5$.
Our recent results show that this might come from the nonlinear response of microscope cameras and should be corrected by calibrating concentrations \cite{ConcCalib}.
For thinner cells the concentration data are much noisier, however it is visible that smearing is smaller and slower.

The interface formation is not perfect even in microfluidics.
As can be seen in Fig.\ref{FIG:MFdiff}, already the initial interface (at small $y$) is smeared (a) or creates an optical effect of bright and dark accents (b)\&(c). 
This makes analysis for concentration profiles at small $y$, i.e. short times $t$, impossible.
Therefore, concentration profiles for earlier time $t$ in Fig.\ref{FIG:ConcPlots}(a)-(c) are already rather smeared.

For comparison, concentration profiles from numerical results are shown in Fig.\ref{FIG:ConcPlots}(d)-(e).
They are chosen so that $Ra_g$ are as similar as possible to the corresponding $Ra_g$ of experimental concentration profiles in (a)-(c). 
In addition, $x$ axis in Fig.\ref{FIG:ConcPlots}(a)-(c) is chosen to agree with Fig.\ref{FIG:ConcPlots}(d)-(f) and differences between times $t$ are equal, if compared in dimensionless units (scaling factors $h/2$ for distance and $h^2/4D$ for time).
Qualitative agreement can be seen.

\begin{figure}
	\centering
		\includegraphics[width=\columnwidth]{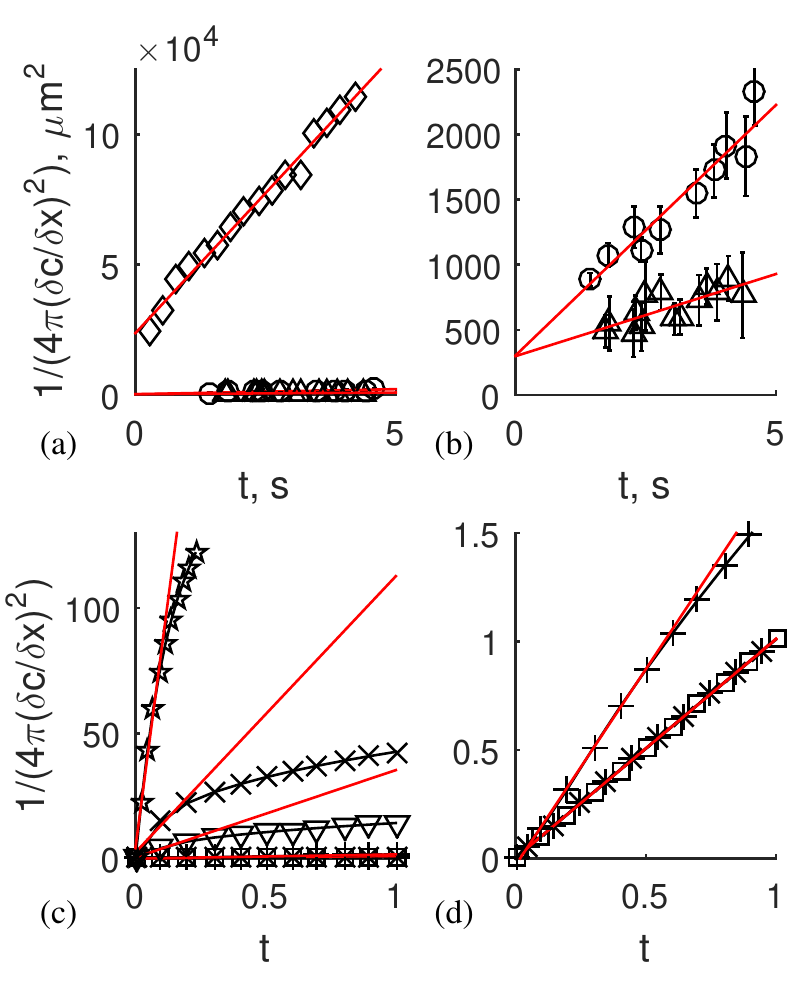}
	\caption{Diffusive behavior of interface smearing. Experimental results in (a) and (b): $\Diamond$ $h=130~\mu$m ($Ra_g\approx13500$), $\bigcirc$ $h=50~\mu$m ($Ra_g\approx900$) and $\bigtriangleup$ $h=25~\mu$m ($Ra_g\approx110$). Numerical results in (c) and (d): $\star$ $Ra_g=15000$, $\times$ $Ra_g=3000$, $\bigtriangledown$ $Ra_g=1000$, $+$ $Ra_g=100$, $\square$ $Ra_g=1$ and $*$ $Ra_g=0.001$. Red lines indicate linear fits.}
	\label{FIG:Deff}
\end{figure}

We use the concentration profiles to characterize dynamics, check for diffusive behavior in concentration smearing and find diffusion coefficients, where applicable.
For that we find $\delta c/\delta x$ for each concentration profile (see fitted slopes that are marked with dotted lines in Fig.\ref{FIG:ConcPlots}).
Following relation \eqref{eq:dcdx}, in Fig.\ref{FIG:Deff} we plot $1/(4\pi\left(\frac{\delta c}{\delta x}\right)^2)$ as a function of time $t$.
The slope we call the effective diffusion coefficient $D_\text{eff}$.

Subplots (a) and (b) show experimental data for the three different cell thicknesses $h$.
For $h=130~\mu$m (diamonds in Fig.\ref{FIG:Deff}(a)) the interface smearing grows linearly with time and is much faster than for two thinner cells.
Also results for $h=50~\mu$m (circles in Fig.\ref{FIG:Deff}(b)) and $h=25~\mu$m (upward pointing triangles in Fig.\ref{FIG:Deff}(b)) indicate linear behavior, while the growth is slower.
The slopes for all series are fitted with linear curves (red lines), which give the effective diffusion coefficients $D_\text{eff}=16.7\cdot10^{-5}$~cm$^2\cdot$s$^{-1}$ for $h=130~\mu$m, $D_\text{eff}=0.38\cdot10^{-5}$~cm$^2\cdot$s$^{-1}$ for $h=50~\mu$m and $D_\text{eff}=0.13\cdot10^{-5}$~cm$^2\cdot$s$^{-1}$ for $h=25~\mu$m.

Subplots (c) and (d) in Fig.\ref{FIG:Deff} show numerical results for multiple $Ra_g$. 
As seen also in concentration profile dynamics, larger $Ra_g$ results in faster interface smearing.
Compared to experimental data, non-linear regimes can be seen for large $t$.
However, they correspond to much longer times than in experiments.
For example, $t=5$~s for $h=130~\mu$m ($Ra_g=13500$) is $t=0.03$ in dimensionless units.
For a similar $Ra_g=15000$ (stars in Fig.\ref{FIG:Deff}(c)), the linear regime is up to $t\approx0.1$ and corresponding is $D_\text{eff}=790$.
Decreasing $Ra_g$ results in a linear smearing and a constant slope, as visible in Fig.\ref{FIG:Deff}(d), where $Ra_g=1$ (squares) and $Ra_g=0.001$ (asterisks) overlap.
This corresponds to normal diffusion $D_\text{eff}=D_\text{eff}/D_0=1$, as diffusion coefficient in dimensionless units is $D_0=1$. 
Using normalized effective diffusion coefficient $D_\text{eff}/D_0$ allows to automatically compare numerical and experimental results, without converting units. 

\begin{figure}
	\centering
		\includegraphics[width=\columnwidth]{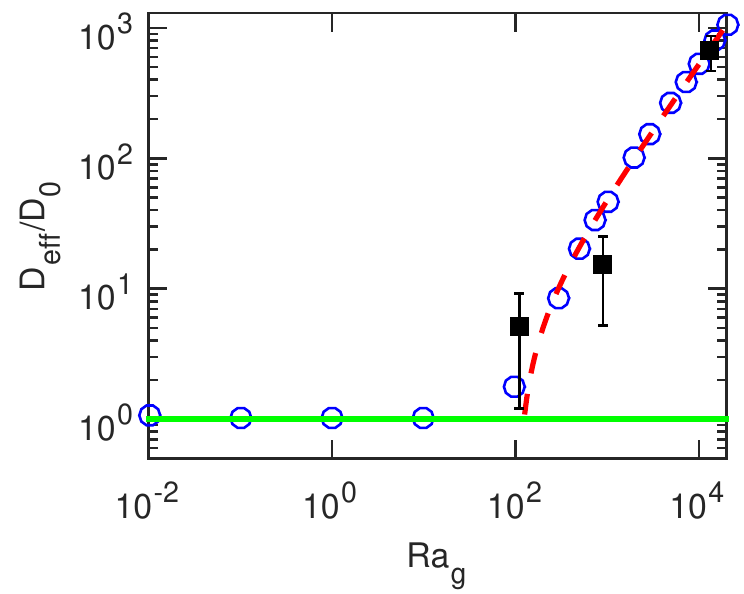}
	\caption{Normalized effective diffusion coefficient $D_{\text{eff}}/D_0$ as a function of gravitational Rayleigh number $Ra_g$. Numerical simulation results displayed with blue circles. Green line shows normal diffusion $D_{\text{eff}}/D_0=1$. Dashed red line is a linear fit. Black squares with errorbars are experimental data.}
	\label{FIG:RaDeff}
\end{figure}

To evaluate the interface smearing dependence on gravitational Rayleigh number $Ra_g$, we plot numerical results for $D_\text{eff}/D_0$ as a function of $Ra_g$.
This is shown in Fig.\ref{FIG:RaDeff} using log--log coordinates for clearer visibility, as the investigated region includes several orders of magnitude.
Two different dependencies can be observed.
For small $Ra_g<100$, the effective diffusion coefficient $D_\text{eff}$ is equal to real diffusion coefficient of particles $D_0$.
For larger $Ra_g>100$, the effective diffusion coefficient $D_\text{eff}$ grows linearly with gravitational Rayleigh number $Ra_g$, following relation:
\begin{equation}
    D_{\text{eff}}/D_0=0.053\cdot(Ra_g-Ra_g^c),
\end{equation}
where $Ra_g^c=105$ is critical gravitational Rayleigh number.
More details can be found in \cite{KitenbergsThesis}.

For comparison, we calculate $D_\text{eff}/D_0$ for experimental data.
Using $D_0=2.5\cdot10^{-7}$~cm$^2\cdot$s$^{-1}$, we get $D_\text{eff}/D_0=670$ for $h=130~\mu$m, $D_\text{eff}/D_0=15.2$ for $h=50~\mu$m and $D_\text{eff}/D_0=5.2$ for $h=25~\mu$m.
These points are shown with black squares in Fig.\ref{FIG:RaDeff}.
Errorbars are calculated from uncertainties in fits of $D_\text{eff}$.
One can see a reasonably good agreement.
This confirms the gravitational influence on the magnetic micro--convection.

Similar gravity-induced interface reorientation between two liquids of different densities in microfluidics has been previously observed experimentally in \cite{InterfaceGravity,Bennet}.
It has also been investigated numerically \cite{GravitySimulation}.
However, these studies have neglected the diffusion of particles.
In this case the water based magnetic fluid and water interface, together with the magnitude of colloidal diffusion and density difference, form particular conditions, where intermediate effects can be observed.
This has allowed us to develop a theoretical model that predicts the gravitational influence and can be useful for development of future applications. 

A well known effect, which also uses an effective diffusion coefficient in its description, is the Taylor-Aris (TA) dispersion \cite{TaylorD}.
It is observed for a pressure driven flow in a thin channel, which distorts the concentration faster than diffusion due to the parabolic flow profile.
The effective diffusion coefficient is in the form $D_\text{eff}^{TA}=D_0\cdot[1+\beta {Pe}^2]$, where ${Pe}=vh/D_0$ is the dimensionless P\'{e}clet number, but $\beta$ is a geometric factor and for a flow between two parallel plates $\beta=1/210$ \cite{TA,TAMF}.
Although the TA dispersion is relevant to the microfluidic channel size and velocities we use, it does not affect the concentration distribution we capture in images close to the beginning of the joint channel.
Instead, it distorts the concentrations further along the channel, while the gravitational convection is quicker at the very beginning.
If we compare the effective diffusion description, we first have to relate the gravitational Rayleigh number $Ra_g$ that we use with the P\'{e}clet number ${Pe}$.
Following the definition of gravity induced velocity in \S\ref{sec:gravity}, one can show that $Ra_g~Pe$, where the proportionality constant depends on the scaling used.
From this we can directly compare $D_\text{eff}/D_0$ in our results with $D_\text{eff}^{TA}/D_0$ and observe that the power of dimensionless quantities is different.
For TA dispersion the effective diffusion coefficient growth with the square of ${Pe}$, while for our case the increase is linear with $Ra_g$.
Hence, the gravity induced convection is quite different from the TA dispersion.
Indeed, the actual concentration mixing in TA dispersion is very different than the convective motion that only visually resembles diffusion.

Changing the thickness of the cell $h$ also allows to expand the verification of the theoretical model of the magnetic micro--convection \cite{JFM2}, introduced in \S\ref{sec:mcv}.
One of the predictions is the change of finger size.
Characteristic fingering patterns of instability for the three cells are shown in Fig.\ref{FIG:Pattern}.
One can observe that for $h=50~\mu$m the characteristic finger size has reduced to $\lambda=50~\mu$m (see Fig.\ref{FIG:Pattern}(b)), exactly as predicted previously
However, for the thinnest cell $h=25~\mu$m the observed finger size is $\lambda=35~\mu$m and is slightly larger than thickness $h$. 
The difference might come from the fact that this image is made for a slowly moving interface and the initial smearing varies along it.

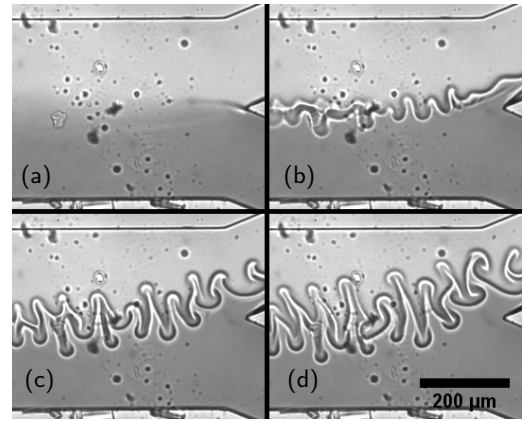
\begin{figure}
	\centering
		\includegraphics[width=0.8\columnwidth]{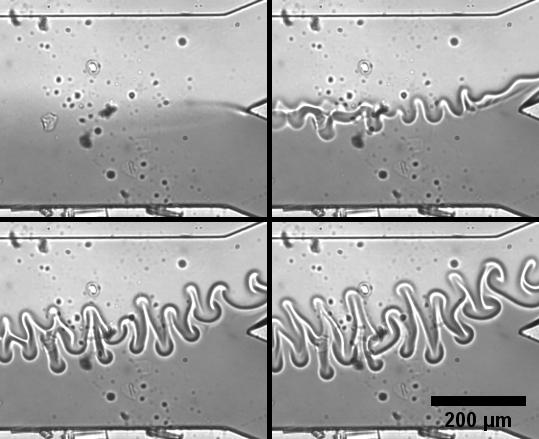}
	\begin{picture}(0,0)
	\unitlength=\columnwidth
	\textcolor{black}{\put(-0.8,0.37){(a)}}
	\textcolor{black}{\put(-0.4,0.37){(b)}}
	\textcolor{black}{\put(-0.82,0.05){(c)}}
	\textcolor{black}{\put(-0.42,0.05){(d)}}
	\end{picture}
	\caption{Dynamics of magnetic micro--convection in microfluidics with $Q=0.1~\mu$L/min at $H=92$~Oe for a cell of thickness $h=50~\mu$m. Corresponding times are (a) $0$~s, (b) $0.2$~s, (c) $0.4$~s, (d) $0.6$~s}
	\label{FIG:MCVMF}
\end{figure}

Another parameter to verify is the change in the critical magnetic field $H_c$ for different thicknesses.
As for the thinner cells we are unable to have an interface with no flow conditions, we look for a critical magnetic field $H_c$ at different flow rates $Q$, as was done in \cite{EPJE}.
This allows to extrapolate the critical magnetic field at zero flow rate.
This means trying multiple magnetic fields $H$ for each flow rate $Q$, until no more instability is observed on the interface.
A characteristic image series of magnetic micro--convection dynamics is shown in Fig.\ref{FIG:MCVMF}, where a situation with magnetic field rather far from the critical field can be seen.
A clear fingering instability appears and develops.

\begin{figure}
	\centering
		\includegraphics{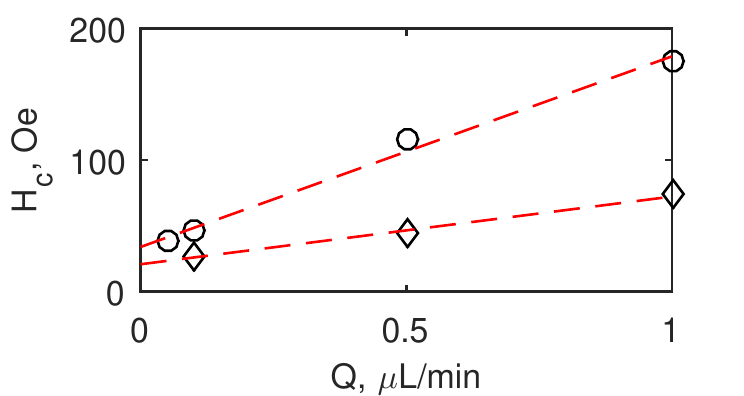}
	\caption{The critical magnetic field $H_c$ dependence on the flow rate $Q$ for two different cell thicknesses $h=50~\mu$m (diamonds) and $h=25~\mu$m (circles). Red dashed lines are linear fits.}
	\label{FIG:Hcrit}
\end{figure}

Results of critical magnetic fields $H_c$ for the two thinner cells and several flow rates $Q$ are summarized in Fig.\ref{FIG:Hcrit}.
Similar to results in \cite{EPJE}, critical magnetic field shows a linear dependence on flow rate $Q$.
We fit the data with linear curves (red dashed lines in Fig.\ref{FIG:Hcrit}) and extrapolate critical magnetic fields at zero flow rate, $H_c=21$~Oe for $h=50~\mu$m $H_c=34$~Oe for $h=25~\mu$m.
From measurements in \cite{JFM2}, the critical magnetic field for $h=130~\mu$m was found to be $H_c=19$~Oe.

Using critical magnetic Rayleigh number $Ra_m^c\approx6$ and the measured effective diffusion coefficients $D_\text{eff}$ for each cell thickness $h$, we can calculate the theoretical critical field $H_c=\sqrt{12\eta D Ra_m^{\text{crit}}}/(\chi\cdot h)$.
This gives $H_c=53$~Oe for $h=130~\mu$m, $H_c=21$~Oe for $h=50~\mu$m and $H_c=24$~Oe for $h=25~\mu$m.
Experimental measurements agree well only for $h=50~\mu$m, while for $h=130~\mu$m the experimentally measured value is more than two times smaller.
However, the calculated value is close to the characteristic field where transition between straight and bent fingers appear $H\approx40$~Oe \cite{JFM2}.
This might indicate that the reason for appearance of straight fingers might not come from the magnetic micro--convection.
The differences for $h=25~\mu$m are smaller and might come from the flow fluctuations.
At the moment our experimental system often experiences pressure oscillations, as typical for small microfluidics channels. 
Hence, taking into account these clarifications, also critical field observations are consistent with the model predictions in \cite{JFM2}.

\section{Conclusions}
We have investigated the interplay of magnetic, diffusive and gravitational effects on the magnetic micro-convection.
A small density difference between miscible magnetic and non-magnetic fluids is sufficient to form a gravity induced convection within a thin cell.
A theoretical model, depending on a single dimensionless gravitational Rayleigh number $Ra_g$, explains the phenomenon.
We verify it with numerical simulations and experiments with different cell thicknesses.
Characteristic interface smearing recalls diffusive behavior and parasitic gravitational convection disappears in a cell that is thin enough.
Although, the conditions look similar to the Taylor-Aris dispersion, these phenomena are quite different.
When gravity is excluded, magnetic micro-convection can still be described by previously developed Brinkman model. 

In addition, gravitational Rayleigh number $Ra_g$ can be used to estimate potential gravitational influence on any colloidal system in microfluidics.
This can be helpful for various applications.

\section*{Aknowledgements}
Authors are thankful to colleagues from PHENIX lab at the Sorbonne University: R.Perzynski and E.Dubois for fruitful discussions and D.Talbot for magnetic fluid.

G.K. acknowledges support from PostDocLatvia grant No. 1.1.1.2/VIAA/1/16/197. Authors are thankful to the French-Latvian bilateral program Osmose project FluMaMi (n$^{\circ}$40033SJ; LV-FR/2019/5).

\printcredits

%% Loading bibliography style file
\bibliographystyle{model1-num-names}

% Loading bibliography database
\bibliography{cas-refs}

%\vskip3pt
\end{document}